\documentclass[onecolumn,draftclsnofoot,12pt]{IEEEtran}
\usepackage[T1]{fontenc}
\usepackage{textcomp}
\usepackage{graphicx}
\usepackage[unicode=true,
 bookmarks=true,bookmarksnumbered=true,bookmarksopen=true,bookmarksopenlevel=1,
 breaklinks=false,pdfborder={0 0 0},pdfborderstyle={},backref=false,colorlinks=false]
 {hyperref}
\hypersetup{pdftitle={Your Title},
 pdfauthor={Your Name},
 pdfpagelayout=OneColumn, pdfnewwindow=true, pdfstartview=XYZ, plainpages=false}
\usepackage{breakurl}

\makeatletter
 \let\oldforeign@language\foreign@language
 \DeclareRobustCommand{\foreign@language}[1]{%
   \lowercase{\oldforeign@language{#1}}}

\usepackage[caption=false,font=footnotesize]{subfig}

\@ifundefined{showcaptionsetup}{}{%
 \PassOptionsToPackage{caption=false}{subfig}}
\usepackage{subfig}
\makeatother

\begin{document}

\title{5G Converged Cell-less Communications in Smart Cities}

\author{Tao\ Han,\ \IEEEmembership{Member,\ IEEE}, Xiaohu\ Ge,\ \IEEEmembership{Senior\ Member,\ IEEE},
Lijun Wang,\ \IEEEmembership{Member,\ IEEE}, Kyung\ Sup\ Kwak,\ \IEEEmembership{Member,\ IEEE},
Yujie\ Han, and Xiong\ Liu \thanks{Accepted by IEEE Communications Magazine SI on ``Enabling Mobile
and Wireless Technologies for Smart Cities''. The corresponding author
is Xiaohu Ge.}\thanks{Tao Han, Xiaohu Ge, Yujie Han, and Xiong Liu are with the School of
Electronic Information and Communications, Huazhong University of
Science and Technology, Wuhan, 430074 P.R. China, e-mail: \{hantao,
xhge, hanyujie, m201571774\}@hust.edu.cn.}\thanks{Lijun Wang is with the School of Electronic Information, Wuhan University,
Wuhan, 430072 P.R. China, and the Faculty of Information Science and
Technology, Wenhua College, Wuhan, 430074 P.R. China, e-mail: wanglj22@163.com.}\thanks{Kyung Sup Kwak is an Inha Hanlim Fellow Professor with Department
of Information and Communication, Inha university, Incheon, Korea,
email: kskwak@inha.ac.kr.}\thanks{Copyright (c) 2016 IEEE. Personal use of this material is permitted.
However, permission to use this material for any other purposes must
be obtained from the IEEE by sending a request to pubs-permissions@ieee.org.}\thanks{Digital Object Identifier: 10.1109/MCOM.2017.1600256CM}}

\markboth{IEEE Communications Magazine, vol. 55, no. 3, Mar. 2017.}{}
\maketitle
\begin{abstract}
Ubiquitous information service converged by different types of heterogeneous
networks is one of fundamental functions for smart cities. Considering
the deployment of 5G ultra-dense wireless networks, the 5G converged
cell-less communication networks are proposed to support mobile terminals
in smart cities. To break obstacles of heterogeneous wireless networks,
the 5G converged cell-less communication network is vertically converged
in different tiers of heterogeneous wireless networks and horizontally
converged in celled architectures of base stations/access points.
Moreover, the software defined network controllers are configured
to manage the traffic scheduling and resource allocation in 5G converged
cell-less communication networks. Simulation results indicate the
coverage probability and the energy saving at both base stations and
mobile terminals are improved by the cooperative grouping scheme in
5G converged cell-less communication networks.
\end{abstract}

\newpage{}

\section{Introduction}

Smart cities are the evolution trends of future cities, which involve
in many aspects of the daily life in cities, such as e-businesses,
intelligent transportation systems, telemedicine, metropolis managements,
security surveillances, logistical managements, social networks, community
services and so on. To brace above services, smart cities have been
employing various wireless communication technologies and networks,
including the bluetooth, ZigBee, radio frequency identification (RFID)
wireless technologies, wireless cellular networks, wireless local
area networks (WLANs), radio broadcasting networks, wireless sensor
networks, body area networks and many others \cite{chen_smart_2016_ACMSpringerMob.Netw.Appl.}.
These wireless communication technologies along with fiber communication
networks and cable networks form the ubiquitous networks for smart
cities. Furthermore, these different types of heterogeneous wireless
networks are expected to support the mobile Internet, Internet of
things (IoT), cloud computing \cite{chen_aiwac:_2015_IEEEWirel.Commun.}
and big data in smart cities.

In future smart cities, the different types of information need to
be smoothly transmitted by different types of heterogeneous wireless
networks with the high data rate and the low energy consumption. In
this case, the simple interconnection with different types of heterogeneous
wireless networks can not support the ubiquitous information services
of future smart cities. Furthermore, the mobile converged network
has been proposed to satisfy the high data rate and the low energy
consumption \cite{han_mobile_2014_IEEEWirel.Commun.}. Compared to
the simple interconnection scheme, a new network architecture needs
to be proposed for the convergence of heterogeneous networks based
on different transmission technologies in smart cities. In general,
there are two levels of heterogeneity of communication networks in
smart cities. One of the levels refers to the different transmission
technologies among Bluetooth, ZigBee, WLAN, millimeter wave, and even
visible light communication (VLC) \cite{lynggaard_deploying_2015_WirelessPersCommun},
while another level of heterogeneity is related to different configurations
and parameters of the same transmission technology, e.g., a heterogeneous
cellular network is consisted of a macro cell tier, a few micro-cell
tiers and femto-cell tiers \cite{cimmino_role_2014_Trans.Emerg.Telecommun.Technol.}.
Conventional communication networks including cellular networks and
WLANs have a distinct characteristic, i.e., there are regional areas
around the base stations (BSs) or the access points (APs), in which
mobile terminals have to access the network via its associated BS
or AP. Such an area can be defined as a ``cell'' associated to a
BS in cellular networks, or just a ``covered area'' associated to
an AP in WLANs. In this article, both of them are named as ``cells''
for the sake of convenience. In conventional heterogeneous cellular
scenarios, a mobile terminal has to handover vertically among heterogeneous
network tiers, or handover horizontally among adjacent cells in the
same tier. As a consequence, it is required to perform complicated
switching and routing algorithms across various types of heterogeneous
networks. By some recent research, Software Defined Network (SDN),
which had been introduced to wired networks many years ago, has been
using in managing complicated mobile networks and performing the traffic
routing in new generation networks \cite{chen_software-defined_2016_},
and SDN is also suitable to manage complicated heterogeneous networks
\cite{sun_intelligent_2015} such as 5G networks in smart cities.
Some researches show that SDN can be improved to manage dynamic links
such as access network links or 5G backhaul links, which are widely
required in coordinated multipoint transmission networks \cite{han_small_2016_IEEESensorsJournal}.
To realize ubiquitous and universal network services in smart cities,
we try to use SDN technology to break the technology gaps and regional
strict in both vertically tier-ed and horizontally celled heterogeneous
networks to support the ubiquitous information services in smart cities.

The most important contribution of this article is to show that a
5G converged cell-less communication scheme is proposed to meet rising
challenges in smart cities, such as heterogeneous wireless transmission
technologies and interference. Numerical results indicate that the
proposed 5G converged cell-less communication network has a better
coverage performance and higher energy efficiency compared with conventional
cellular networks. In Section II the architecture and model of cell-less
communication networks are introduced for smart cities. The performance
of the 5G converged cell-less communications is investigated by evaluating
the performance of coverage and energy efficiency in Section III.
The future challenges of the 5G converged cell-less communications
in smart cities are discussed in Section IV. The conclusion is drawn
in Section V.

\section{Architecture and Model of Cell-less Communications in Smart Cities}

\subsection{From Celled Networks to Cell-less Networks}

There are many challenges for the mobile and wireless communications
in smart cities. Based on the development of 5G communication systems,
some of issues involving with the urban scenarios are list as follows.
\begin{enumerate}
\item The huge demand of data rate causes ultra-densified BSs/APs deployment.
There are three main approaches for 5G communication systems to increase
data rate significantly, i.e., the wider spectrum of millimeter wave
transmission, the more spatial diversity of massive MIMO and the more
spatial density of BSs/Aps. Deploying more BSs/APs can serve more
high-data-rate-demanding users and thereby provide a higher achievable
data rate in terms of per unit area in smart cities. The higher density
of BSs/APs makes the smaller cell coverage as the distance between
BSs/APs is reduced to tens meters for satisfying the high-data-rate-demanding
in smart cities \cite{ge_5g_2016_Accept.AppearIEEEWirel.Commun.}.
\item The movement of mobile terminals in modern metropolitan scenarios
becomes more complex and volatile. In a prosperous city, the movement
of mobile nodes is varied and complicated \cite{giust_distributed_2015_IEEECommun.Mag.}.
In smart cities, a mobile terminal with data transmission can be a
mobile phone carried by a pedestrian, a navigation device installed
on a moving car or a PDA used in a high-speed train. And what is more,
various types of mobile nodes make the mobility situation even harder
to handle. For example, there are completely different communication
requirements between the scenarios of densified RFID labels going
through a gate and a mobile high definition surveillance camera moving
around in a disaster scene. 
\item In urban areas, buildings and trees become obstacles to wireless communications.
The wireless communication channels are very different between indoor
and outdoor environments. The designers of mobile communication systems
have to consider the obvious impact of obstacles, especially for the
millimeter wave wireless transmission in the emerging 5G communication
networks, which is of very short wavelength and hard to diffract in
smart cities.
\end{enumerate}
With regard to above demands to the mobile communications in smart
cities, current heterogeneous networks based on different types of
communication technologies meet many issues for the ubiquitous information
services in smart cities, and some of them are listed as below. 
\begin{enumerate}
\item Issue of network convergence. To overcome problems of the vertical
and horizontal handover and routing across tiers, how to converge
the heterogeneous networks becomes a critical issue. For the prevailing
wireless transmission technologies and communication networks, it
is hard to converge them with each other seamlessly. Instead, they
interconnect with each other, in this way, many issues regarding routing
and protocols remain in heterogeneous wireless networks.
\item Issue of load balancing in celled networks. Along with the decreasing
of the cell size, the traffic loads of cells get more and more unbalanced.
Moreover, the traffic load of smart cities is obviously fluctuated
over space and time domains. The fluctuation of traffic load in space
domain is caused by the stochastic spatial distribution of communication
nodes in smart cities, e.g., the data centers of smart cities are
stochastic distributed in different places. The fluctuation of traffic
load in the time domain is created by the mobility of terminals scheduled
by the work and life in smart cities \cite{ge_user_2016_JSAC}. For
the improving of signal-to-interference-plus-noise ratio (SINR) in
wireless communications, the sizes of cells in mobile networks get
smaller to gain higher date rate matching the terminal\textquoteright s
higher data rate demand. As we know, the bigger size of a cell can
smooth the random fluctuation in the space domain. When the size of
a cell is getting small in 5G networks, the traffic load balance issue
emerges for smart cities. 
\item Issue of handover. When the cell size is reduced to tens meters in
5G cellular networks, the quickly moving terminals lead to frequent
handovers in 5G cellular networks and additional latency is inevitable
for wireless communications. When the handover occurs between different
types of heterogeneous wireless networks, a large amount of overhead
in wireless networks will decrease the data exchanging efficiency.
\item Issue of interference. In an interference-limited conventional cellular
network, the increase of BS/AP density doesn\textquoteright t lead
to the increase of the average interference indicator \cite{dhillon_modeling_2012_IEEEJ.Sel.AreasCommun.}.
However, the densified BSs/APs under complicated electromagnetic environments
in smart cities may face the highly correlative interference or noise
and hence the performance of some adjacent BSs/APs drops significantly
\cite{han_interference_2015_Mob.Netw.Appl.}. It is an important concern
to eliminate spatially correlative interference in the dense wireless
networks of smart cities.
\end{enumerate}
From what we discuss above, deploying conventional cellular network
can not solve above issues and satisfy the ubiquitous information
services in smart cities. To solve these problems caused by heterogeneity
of networks and ultra-density of BSs/APs, we propose to use converged
\textquotedblleft cell-less\textquotedblright{} communication networks
instead of \textquotedblleft celled\textquotedblright{} networks to
support the mobile users in smart cities.

\begin{figure}[tbh]
\begin{centering}
\subfloat[Conventional cellular network]{\begin{centering}
\includegraphics[width=0.4\paperwidth]{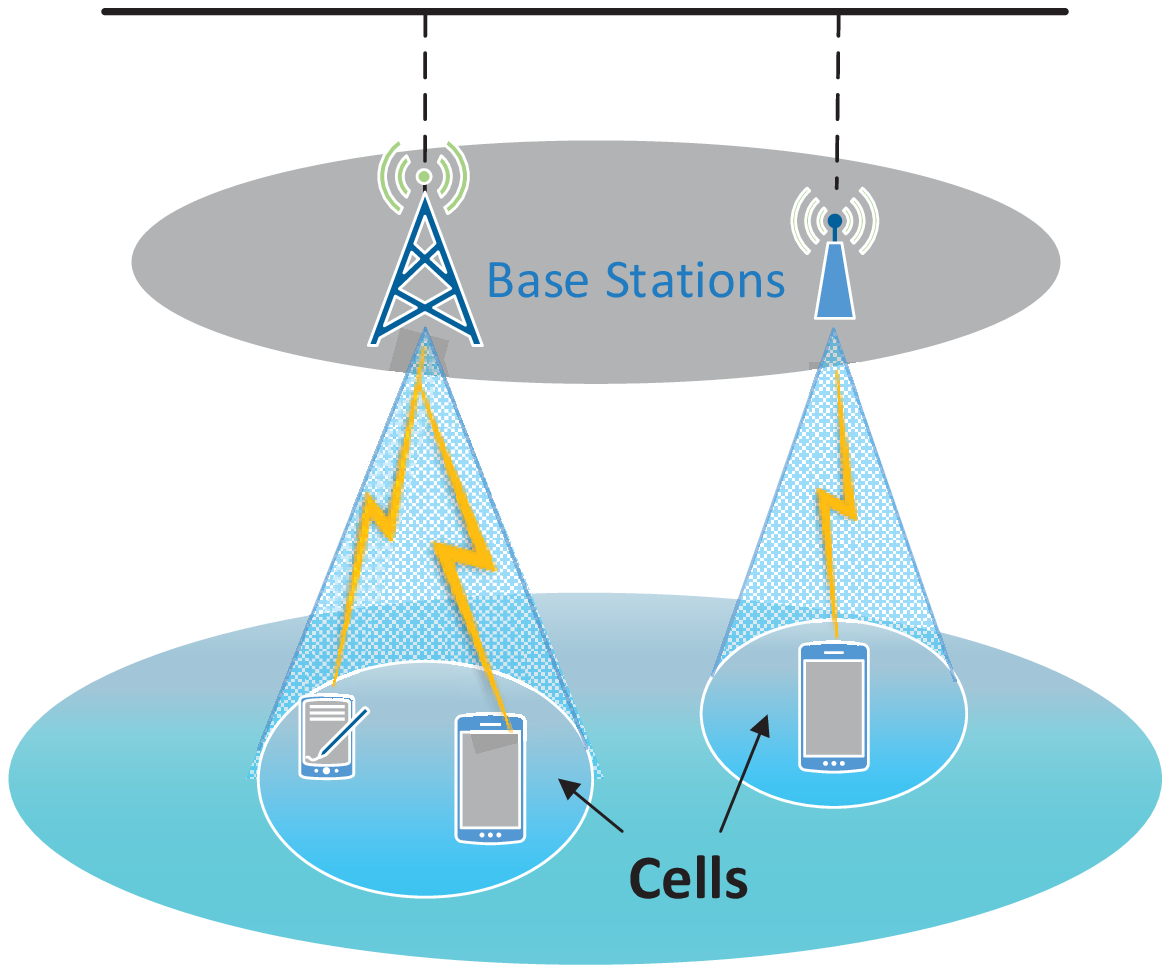}
\par\end{centering}
}\subfloat[Cell-less network]{\begin{centering}
\includegraphics[width=0.4\paperwidth]{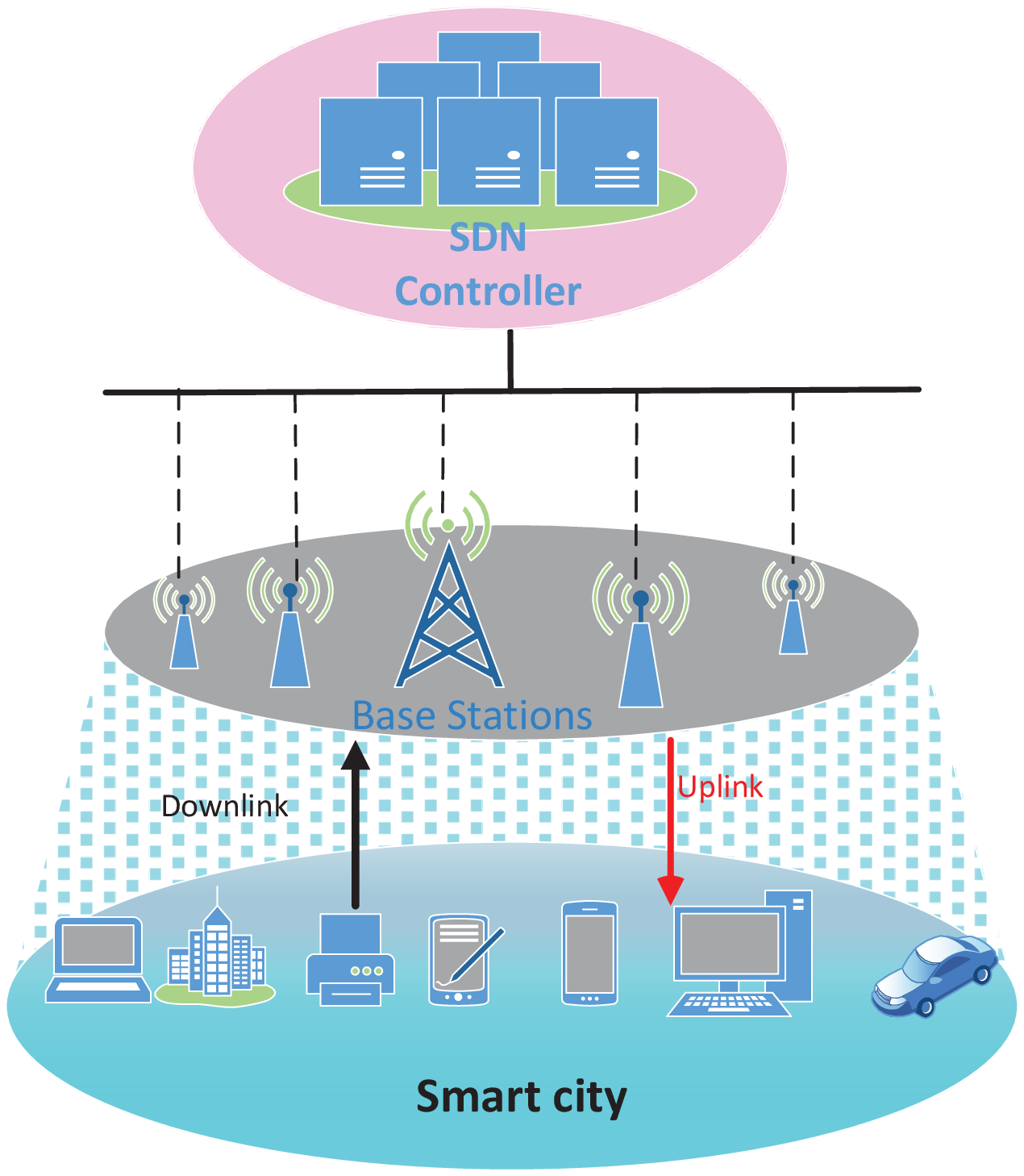}
\par\end{centering}
}
\par\end{centering}
\caption{From a conventional cellular network to a cell-less network\label{fig:Celled_Cell-less}}
\end{figure}

Fig. \ref{fig:Celled_Cell-less} illustrates a conventional cellular
network and a cell-less network in urban scenarios. As shown in the
figure, a mobile terminal in the conventional cellular network always
associates to one and only one BS/AP, while a terminal in a cell-less
network doesn\textquoteright t associate any BS/AP. In this case,
the terminal in a cell-less network can flexibly communicate with
one or more BSs/APs if necessary. In the following parts of the article,
we explain the architecture and transmission model of the cell-less
communication networks.

\subsection{Architecture of Cell-less Communication Networks in Smart Cities}

\begin{figure}[tbh]
\begin{centering}
\includegraphics[width=0.9\textwidth]{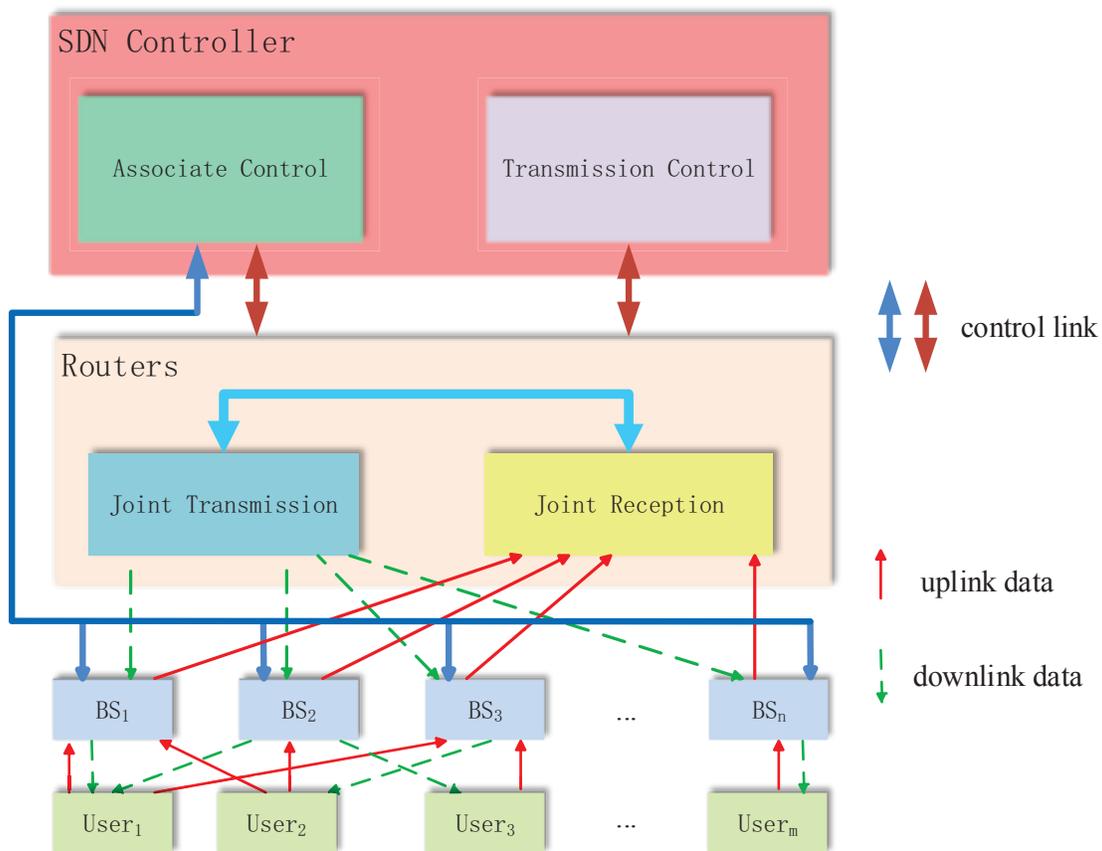}
\par\end{centering}
\caption{Cell-less association relations and data transmission are under the
control of the SDN controller\label{fig:Arch_Cell-less}}
\end{figure}

To match the requirements of huge data rate, ultra-density, high mobility
and low energy consumption of wireless networks in smart cities, the
5G converged cell-less communication network is proposed in this paper.
In the novel cell-less scheme shown in Fig. \ref{fig:Arch_Cell-less},
a mobile terminal can choose to access one or more BSs/APs by different
uplinks and downlinks considering wireless channel status and its
demands, or choose not to access any BS/AP when the mobile terminal
is idle. That is, a mobile terminal doesn\textquoteright t associate
with any BSs/APs before it starts to transmit data. In such case,
BSs/APs need not maintain a list of associated mobile terminals and
instead the SDN controller decides which one or more BSs/APs perform
the data transmission for the mobile terminal by the control link
shown in Fig. \ref{fig:Arch_Cell-less}. Moreover, the SDN controller
creates dynamic backhaul links and downlinks/uplinks as well for the
joint transmission or reception group of BSs/APs such that they can
cooperate with other members in the same group to support joint transmission
and reception for a specified mobile terminal. The cell-less scheme
supports that the number of BSs/APs group is adaptively adjusted by
the requirements from the mobile terminal and the wireless channel
status in different environments. Therefore, the overhead caused by
handover is reduced and the coverage probability is guaranteed for
the mobile terminal. Moreover, the traffic load balancing is achieved
by resource schedule in a large scale network, which is performed
by the SDN controller of converged cell-less communication networks.
Furthermore, the traffic load fluctuation in spatial and temporal
domains is decreased for smart cities. In this cell-less scheme, the
SDN controller and core routers form the SDN cloud, in which the control
plane is driven by the cloud computing, while routers and the instantaneous
backhaul links form the data plane in the cloud.

\subsection{Transmission Model of Cell-less Wireless Communications}

\begin{figure}[tbh]
\begin{centering}
\includegraphics[width=0.9\textwidth]{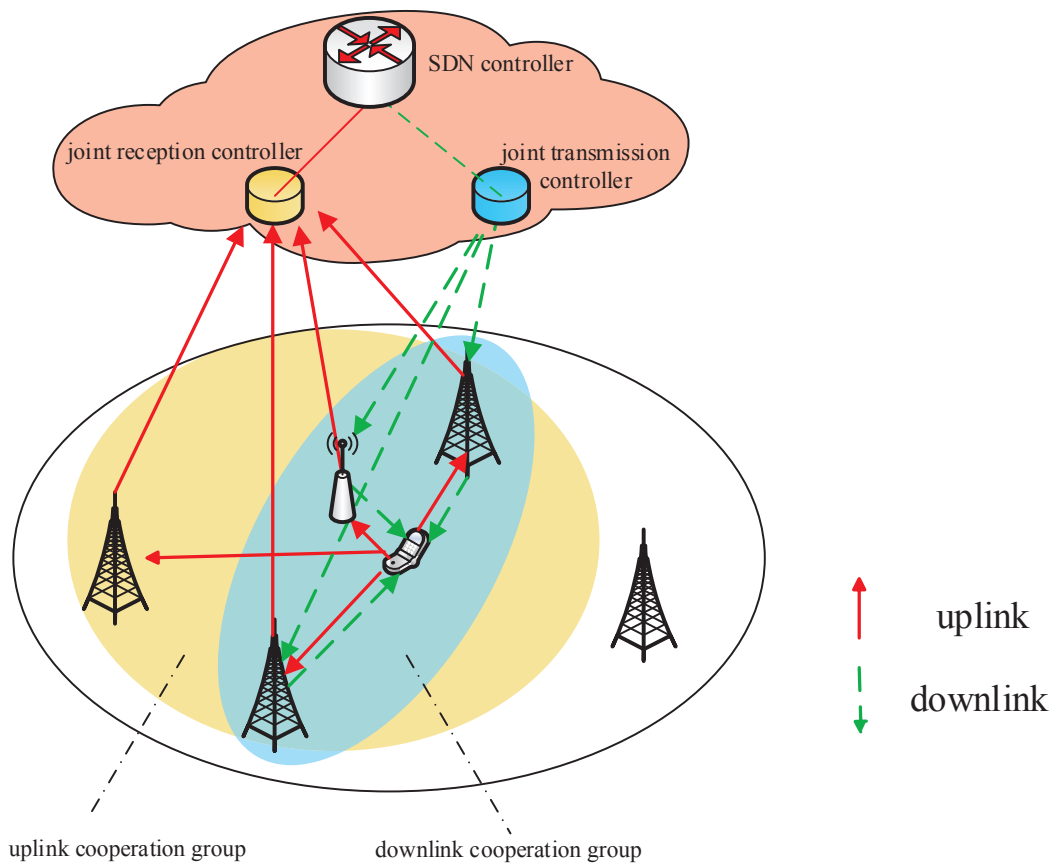}
\par\end{centering}
\caption{Transmission model of a cell-less network\label{fig:UL_DL}}
\end{figure}

To implement the un-associated transmission between BSs/APs and mobile
terminals in cell-less wireless communications, it is necessary to
change the access method. Mobile terminals update their locations
and channel status around them to the SDN cloud in case the communication
to BSs/APs is necessary. As shown in Fig. \ref{fig:UL_DL}, a mobile
terminal transmits the data by broadcasting when it wants to send
the uplink data. Nearby BSs/APs receive the data then forward the
data to the joint reception controller in the cloud where the data
transmitted from the mobile terminal are jointly decoded. When there
are data to be sent to a specified mobile terminal, the SDN controller
in the cloud decide which one or more of the BSs are chosen to form
a cooperative group to perform downlink joint transmission considering
the location and channel status around the terminal.

Compared with the conventional celled networks, the 5G converged cell-less
communication networks have many advantages listed as below.
\begin{enumerate}
\item Seamless convergence of heterogeneous networks. Adopting not only
inter-connection but also data convergence in terms of transmission
environments and user requirements, the cell-less communication networks
provide a compelling mechanism to fulfill convergence across tiers
and combine their respective advantages as well. 
\item Superior traffic load management. By cooperation of dynamically grouped
BSs/APs in the cell-less scheme, the cell-less communication network
can allocate traffic load to BSs/APs under the schedule of the SDN
controller. 
\item Avoiding frequently handovers. In a converged cell-less communication
network, a mobile terminal need not associate to any fixed BS. Hence,
the frequent handovers between cells are avoided which conduces to
the decrease of outage and latency in converged cell-less communication
networks. 
\item Improving coverage and energy efficiency. When the fixed cell association
scheme is given up in heterogeneous wireless networks, the converged
cell-less communication networks reorganize the association scheme
between the mobile terminals and the BSs/APs in terms of the requirements
from users and wireless environments in smart cities. When the flexible
grouped cooperative communication is performed, the improved coverage
probability is expected for a mobile terminal in converged cell-less
communication networks. Moreover, when the suitable BSs/APs are selected
for joint transmission and reception, the energy consumption is also
expected to be optimized. 
\end{enumerate}

\section{Coverage and Energy Efficiency of Converged Cell-less Communication
Networks}

\subsection{BS Grouping Scheme for Converged Cell-less Communications Networks}

How to form the cooperative group of BSs/APs is a critical issue in
converged cell-less communication networks. In generally, the grouping
scheme depends on the spatial distribution of BSs/APs and the wireless
channel environments in smart cities. The basic criteria of cooperative
BSs/APs grouping are suggested as below.
\begin{enumerate}
\item Criterion of simplicity. Considering the ultra-dense deployment of
BSs/APs in smart cities, it is suggested that each station/point serves
only one mobile terminal every time if any possible, but one BS/AP
is allowed to serve more than one mobile terminal in case there may
be congestion in high traffic load scenarios.
\item Criterion of economy. As less as possible BSs/APs are selected to
form cooperative group, given that the group of BSs/APs meet the user
data rate demand.
\item Criterion of uniformity between grouping for uplinks and downlinks.
The SDN controller will always try to keep the same group for both
the uplink and downlink transmission if possible. However, the BSs/APs
of the cooperative groups for uplinks and downlinks can be different
from each other, especially when the mobile terminal move quickly.
\item Employing backhaul multicast capability if any possible. In order
to reduce the backhaul overhead, the downlink data is transmitted
to the cooperative group by multicast methods if possible.
\item Mobility predicting for adjacent mobile terminals. When the BSs/APs
are grouped for an active mobile terminals, it is necessary to acquire
the distribution of adjacent active and inactive mobile terminals
and predict the transmission and reception actions of adjacent active
mobile terminals. Furthermore, the size of the cooperative group is
optimized to avoid the traffic congestion in the hot point.
\item Pre-grouping of BSs/APs. Considering that it may be frequently grouped
for the cooperative BSs/APs in high traffic load scenarios, a pre-grouping
scheme is required to accelerate the grouping speed of cooperative
BSs/APs. The pre-grouping scheme is designed by evaluating the recent
cooperative grouping result to reduce the computational complexity
of cooperative grouping algorithm.
\end{enumerate}
Generally speaking, for indoor scenario where terminals hardly move,
the cooperative BS/AP group needn't be adjusted frequently, while
frequent adjusts must be done for quickly moving terminals outdoors.
To maintain the quality of communication for fast-moving terminals,
more BS/AP candidates are beneficial for BS/AP grouping in downlink
transmission. Moreover, to maintain a consistent quality of communication,
the grouping size should be adjusted when data rate demands of the
users vary.

\subsection{Coverage Probabilities in Converged Cell-less Communication Networks}

As we know, ultra-dense BSs/APs can be deployed to achieve high data
rate in smart cities. Moreover, the coverage of every BS/AP will be
reduced by massive MIMO and millimeter wave communication technologies.
In this case, the cooperative grouping scheme is a reasonable approach
to satisfy the coverage requirement of mobile terminals in smart cities.

\begin{figure}[tbh]
\begin{centering}
\includegraphics{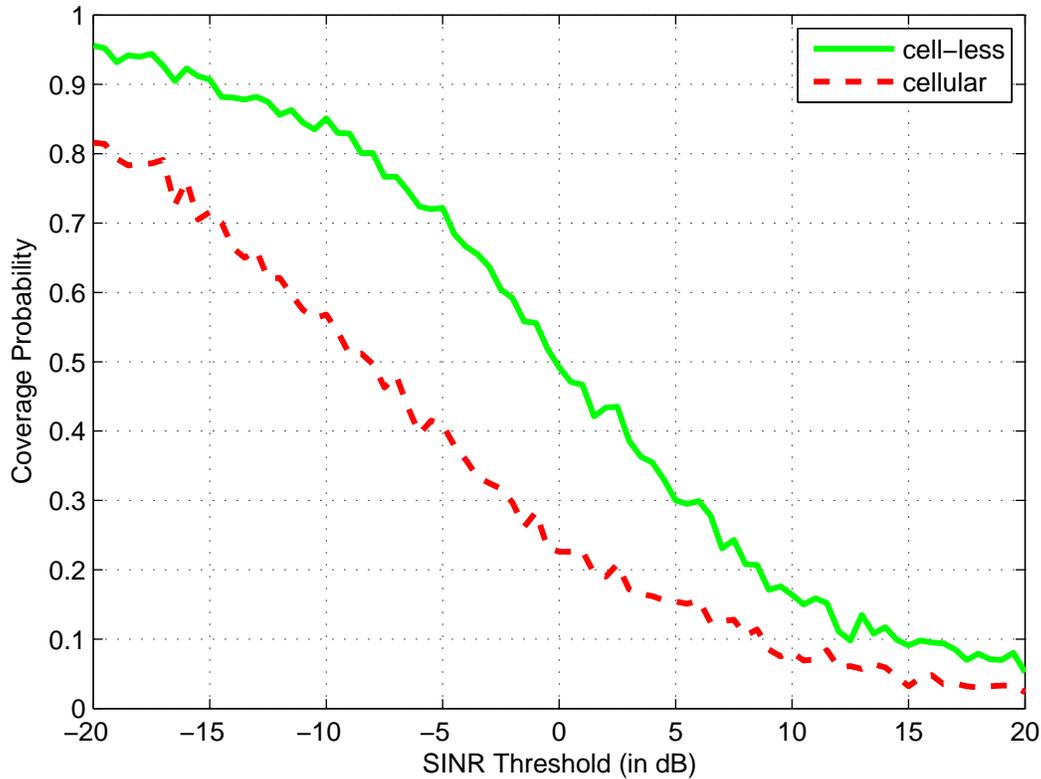}
\par\end{centering}
\caption{Coverage probabilities of cellular and cell-less networks \label{fig:Coverage}}
\end{figure}

Without loss of generality, in our illustrative example, 50 BSs are
deployed into a plane of 50 m \texttimes{} 50 m randomly. Moreover,
30 BSs are configured as members in active cooperative groups. A typical
user is assumed to be located at the central location in the plane.
For the sake of simplicity, the nearest 10 BSs around the typical
user are configured to be candidates for cooperative grouping in converged
cell-less communication networks. The size limit of the cooperative
group is no more than 3 BSs in converged cell-less communication networks.
If there is no idle BS in candidate BSs, the nearest BS around the
typical user is selected to transmit data even this BS is active in
other cooperative group. The coverage probability is analyzed by Monte
Carlo simulations in Fig. \ref{fig:Coverage}. The results indicate
that the coverage probability of the illustrative converged cell-less
communication networks is higher than the coverage probability of
conventional cellular networks when the SINR threshold of the user
terminal is configured as \textminus 15 dB \textasciitilde{} 5 dB.
The reason is that the cooperative group formed by local BSs can significantly
reduce the interference among BSs by converting the interference within
the group into the useful signal.

\subsection{Energy Efficiency in Converged Cell-less Communication Networks}

A large number of BSs/APs are ultra-densely deployed in smart cities.
Hence, there exists the redundancy for BSs/APs when the traffic load
is low in some scenarios, such as the work office in middle nights.
The converged cell-less communication network provides a flexible
BS/AP sleeping scheme to decrease the energy consumption in smart
cities which is controlled by the SDN cloud computing. The detailed
BS/AP sleeping scheme is explained as follows.
\begin{enumerate}
\item A BS/AP can be configured in several states including transferring,
ready, listening and sleeping. When a BS/AP is in the transferring
state, the BS/AP can transmit data to the specified user terminal
or probably quit the active cooperative group due to the dynamic grouping
scheme. After that, the BS/AP turns to ready state.
\item When a large amount of data are transferred, the transmission power
can be dynamic allocated among the members of a cooperative group,
according to the channel status between the user and the BSs/APs.
A deliberate power allocation scheme will make sense to save energy.
\item In low traffic load scenarios, e.g., the work office in midnight,
some of BSs/APs can turn to the sleeping state from ready or listening
state to save as much energy as possible. To guarantee the coverage
probability of converged cell-less communication networks, the active
BSs/APs adaptively increase the coverage area by increasing the transmission
power or grouping more cooperative members if necessary.
\end{enumerate}
\begin{figure}[tbh]
\begin{centering}
\subfloat[Energy saving of BSs\label{fig:sub_EE_BS}]{\begin{centering}
\includegraphics[width=0.32\textheight]{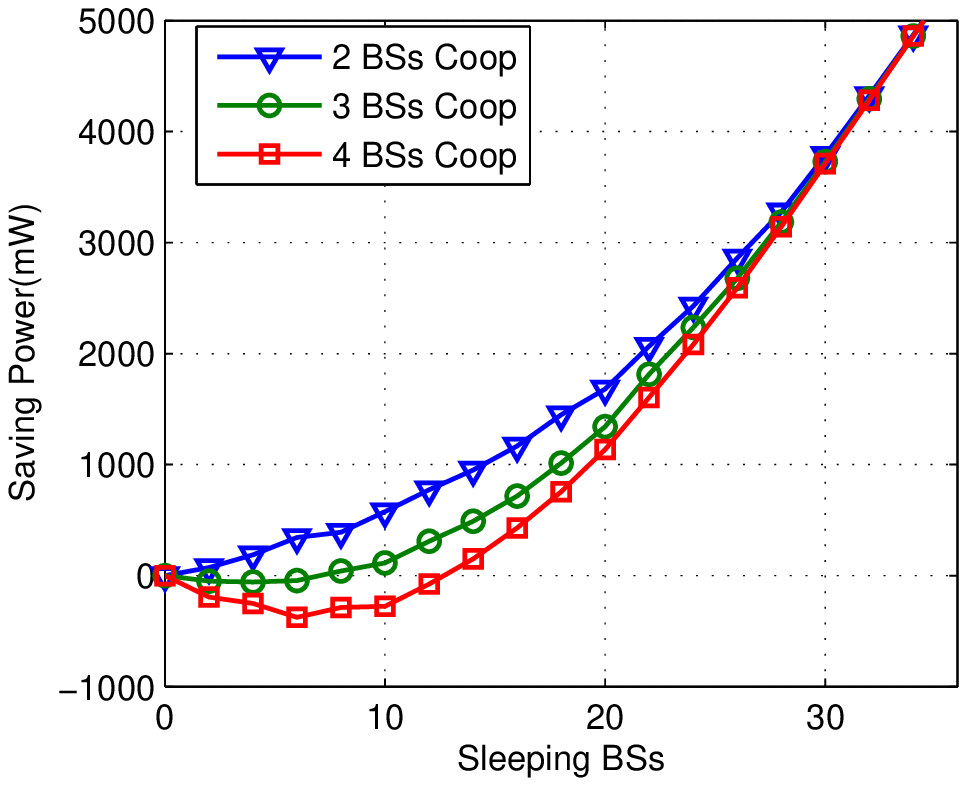}
\par\end{centering}
}\subfloat[Energy saving of mobile terminals\label{fig:sub_EE_MT}]{\begin{centering}
\includegraphics[width=0.48\textwidth]{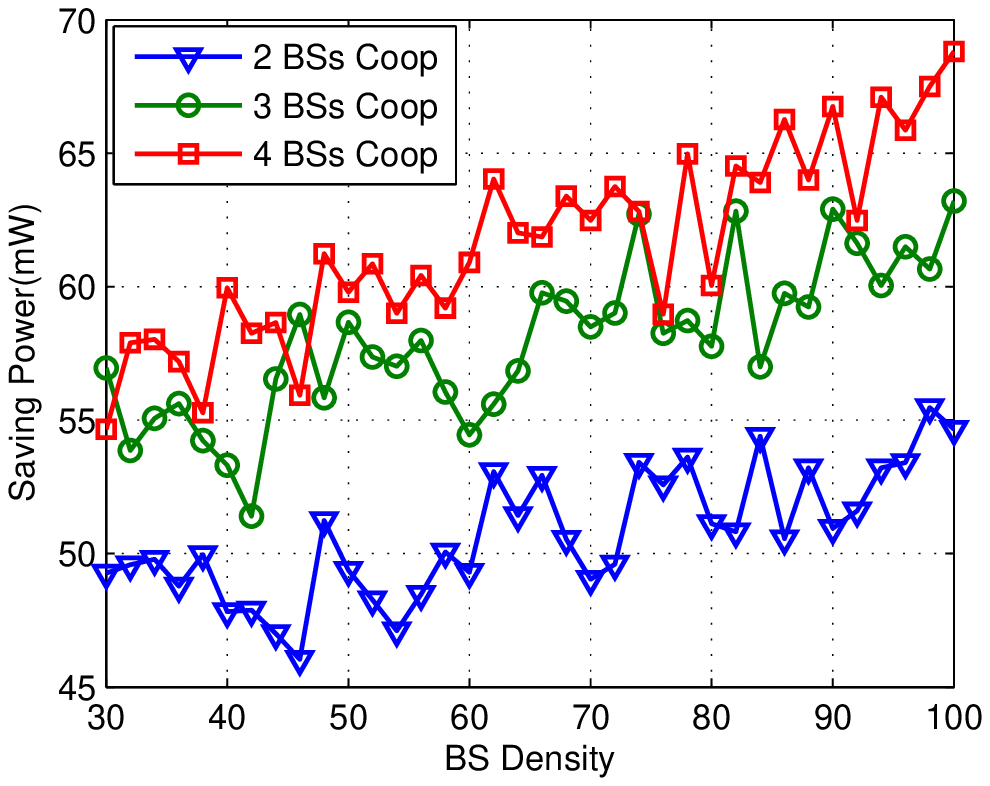}
\par\end{centering}
}
\par\end{centering}
\caption{Energy saving in cell-less networks\label{fig:EE}}
\end{figure}

Utilizing the same illustrative simulation scenario in the Fig. \ref{fig:Coverage},
20 active BSs are selected in random transferring or ready state and
the other BSs are configured into the sleeping state in converged
cell-less communication networks. When a BS is sleeping, its neighbor
BSs are configured to serve active users to guarantee coverage probability
in this area. Considering small BSs in 5G mobile communication systems,
the consumption power of BSs is configured as 10, 50, 80, 200 mW corresponding
to the sleeping, listening, ready and transferring states, respectively.
The energy saving of converged cell-less communication networks with
respect to the number of sleeping BSs considering different numbers
of cooperative BSs is illustrated in Fig. \ref{fig:EE}(a). Numerical
results indicate that the energy saving of BSs increase with the increase
of the number of sleeping BSs in the illustrative converged cell-less
communication networks. Moreover, the cooperative group of two BSs
achieves the maximum energy saving of BSs compared with the cooperative
group of three and four BSs in converged cell-less communication networks.

Without loss of generality, the original transmission power of mobile
terminal is configured as 100 mW and then the received data rate in
non-joint reception scenario can be obtained by simulation firstly.
When the joint reception scheme is adopted in converged cell-less
communication networks, the mobile terminal can adaptively adjust
the transmission power to acquire the same date rate as non-joint
reception scenario. When the joint reception is configured at uplinks,
the energy saving of mobile terminal is shown in Fig. \ref{fig:EE}(b).
Numerical results show that the energy saving of mobile terminal increases
with the increase of the number of cooperative BSs in the illustrative
converged cell-less communication networks. These results imply that
the converged cell-less communication networks save energy not only
at BSs but also at the mobile terminals. 

\section{Future Challenges for Converged Cell-less Communications in Smart
Cities}

\subsection{Cooperation in Ultra-Densified Wireless Transmitters}

In the future smart cities, there exist not only a larger number of
BSs but also many other WLAN APs and IoT nodes. In this case, these
wireless transmitters are ultra-densely deployed in smart cities.
All these wireless transmitters could be converged to provide information
to users by a cell-less network architecture. The cooperative communication
in ultra-densified wireless transmitters is an attractive solution
for converged cell-less communication networks. However, the association
relationship of the cooperative group can not be fixed in advance
considering heterogeneous wireless transmitters in different wireless
networks. In this article, we discuss the potential grouped solutions
and validate the advantages in the mobile user coverage probability
and the energy saving in smart cities. Anyway, there still exist many
challenges need to be investigated. For instance, how to trade-off
the complex and efficiency metrics in cooperative schemes of converged
cell-less communication networks. When the large data rate is available
for mobile terminals, how to realize the backhaul traffic in converged
cell-less communication networks is a great challenge especially considering
different transmission capacities of heterogeneous wireless networks
\cite{ge_5g_2014_IEEENetw.}. The cooperative backhaul solution is
a potential solution to satisfy requirements of the big data collection
and the environment awareness from smart cities. As a consequence,
the investigation of cooperative backhaul schemes in converged cell-less
communication networks is an emerging issue for future smart cities.

\subsection{Data and Control Information in Smart Cities}

To support the cooperative transmission in converged cell-less communication
networks, a part of control information need to be separated from
the transmission data and then the common data could be easily transmitted
and converged in smart cities. Considering the difference among heterogeneous
wireless networks and the architecture of converged cell-less communication
networks, it is an important challenge to design the compatible protocol
of converged cell-less communication networks. Moreover, the information
and data in smart cities have different priority and security level.
In this case, the control information of transmission data can not
be separated in some specified scenarios of smart cities, for some
reasons such as the personal privacy and the public security. Therefore,
a special scheme maybe need to be included into 5G converged cell-less
communications for smart cities. In technology level, it is also a
troubled problem to execute a single special scheme in all heterogeneous
devices even these devices belong to different owners.

\subsection{Cloud and Cache Computing in Converged Cell-less Communications}

As discussed in this paper, the cell-less communications provide a
flexible solution in coverage and energy efficiency for future smart
cities. The cell-less communications can solve the heterogeneous issues
in physical level. To match the advantages brought by the cell-less
communications, the cloud and cache computing schemes are expected
to collaborate with cell-less communications in smart cities. How
to coverage the cloud, cache and cell-less communication into the
uniform architecture for supporting smart cities is a true challenge
for researchers around the world. One of possible way is explored
to coverage above three architectures by the converged data information.
However, there exist different definitions and understandings of the
data and information in the cloud, cache and cell-less communications
in smart cities. More studies need to be carried out to investigate
the converged cloud, cache and cell-less communications. Consequently,
the smart cities will be a good platform to draw the dream of converged
cloud, cache and cell-less communications \cite{chen_towards_2013_MultimediaToolsandApplications}.

\section{Conclusion}

The information and data generated from different types of heterogeneous
wireless networks are converged to provide the ubiquitous service
in smart cities. To support mobile users in smart cities, the idea
of converged cell-less communication networks is proposed to break
the conventional celled architecture of cellular networks and support
the flexible mobile user association scheme considering the application
requirements and wireless channel status. With the deployment of 5G
ultra-dense wireless networks in smart cities, the cooperative group
communication is designed for the 5G converged cell-less communication
networks. Simulation results indicate that the coverage probability
and the energy saving at BSs and mobile terminals are improved in
5G converged cell-less communication networks. Based on the analysis
and illustrative results, it can be concluded that the converged cell-less
communication scheme is a promising way to match the high demand for
coverage and rate in future smart cities, because of its flexibility
and unitarity. Considering there should be many complicated factors
in the future smart cities, such as the high demand caused by crowded
people, the serious obstacle due to a lot of buildings and the heavy
interference in dense streets, the converged cell-less communication
networks can play a critical role because it can converge different
communication technologies and provide seamless transmission and thus
improve coverage and energy efficiency by reducing unnecessary interference.
With the development of smart cities, 5G networks still need to be
further investigated for solving new challenges in smart cities.

\section*{Acknowledgment}

The corresponding author is Xiaohu Ge. The authors would like to acknowledge
the support from the International Science and Technology Cooperation
Program of China (grants 2015DFG12580 and 2014DFA11640), the National
Natural Science Foundation of China (NSFC) (grants 61471180 and 61210002),
the Hubei Provincial Department of Education Scientific research projects
(No. B2015188), the Fundamental Research Funds for the Central Universities
(HUST grants 2015XJGH011 and 2015MS038), the grant from Wenhua College
(No. 2013Y08), the National Research Foundation of Korea-Grant funded
by the Korean Government (Ministry of Science, ICT and Future Planning)-NRF-2014K1A3A1A20034987,
the EU FP7-PEOPLE-IRSES (Contract/Grant No. 318992 and 610524), and
the EU H2020 project (Grant No. 723227). This research is supported
by the China International Scientific and Technological Cooperation
Base of Green Communications and Networks (No. 2015B01008).

\bibliographystyle{IEEEtran}
\bibliography{mcom2016}

\begin{IEEEbiographynophoto}{Tao Han}
 {[}M'13{]} (hantao@hust.edu.cn) received his Ph.D. degree in information
and communication engineering from Huazhong University of Science
and Technology (HUST), Wuhan, China in December, 2001. He is currently
an associate professor with the School of Electronic Information and
Communications, HUST. His research interests include wireless communications,
multimedia communications, and computer networks. He is currently
serving as an Area Editor for the \emph{EAI Endorsed Transactions
on Cognitive Communications}.
\end{IEEEbiographynophoto}

\begin{IEEEbiographynophoto}{Xiaohu Ge}
 {[}M'09-SM'11{]} (xhge@hust.edu.cn) is currently a full professor
with the School of Electronic Information and Communications at Huazhong
University of Science and Technology (HUST), China and an adjunct
professor with at with the Faculty of Engineering and Information
Technology at University of Technology Sydney (UTS), Australia. He
received his Ph.D. degree in information and communication engineering
from HUST in 2003. He is the director of China International Joint
Research Center of Green Communications and Networking. He has published
more than 110 papers in international journals and conferences. He
served as the general Chair for the 2015 IEEE International Conference
on Green Computing and Communications (IEEE GreenCom). He has served
as an Editor for the \emph{IEEE Transaction on Green Communications
and Networking}, etc.
\end{IEEEbiographynophoto}

\begin{IEEEbiographynophoto}{Lijun Wang}
 {[}M'16{]} is pursuing her Ph.D. degree with Wuhan University, Wuhan,
China. She is currently a associate professor with the Faculty of
Information Science and Technology, Wenhua College, Wuhan, China.
Her research interests include wireless communications, and multimedia
communications.
\end{IEEEbiographynophoto}

\begin{IEEEbiographynophoto}{Kyung Sup Kwak}
 received his Ph.D. degree from the University of California at San
Diego in 1988. He had worked for Hughes Network Systems and IBM Network
Analysis Center, USA, and is now with Inha University, Korea as Inha
Hanlim Fellow professor. He served as the president of Korean Institute
of Information and Communication Sciences in 2006, and the president
of Korea Institute of Intelligent Transport Systems in 2009. His research
interests include mobile communications, and wireless sensor networks
including Nano networks.
\end{IEEEbiographynophoto}

\begin{IEEEbiographynophoto}{Yujie Han}
 received his Bachelor\textquoteright s degree in communication and
information system from Huazhong University of Science and Technology,
Wuhan, China, in 2012, where he is currently working toward his Master\textquoteright s
degree. His research interests include cooperative communication,
stochastic geometry, and heterogeneous networks. 
\end{IEEEbiographynophoto}

\begin{IEEEbiographynophoto}{Xiong Liu}
 received his Bachelor\textquoteright s degree in electronic information
and communication from Huazhong University of Science and Technology,
Wuhan, China, in 2015, where he is currently pursuing his Master\textquoteright s
degree. His research interests include vehicular networks, non-orthogonal
multiple access, and cognitive radio. 
\end{IEEEbiographynophoto}

\end{document}